\documentstyle{article}

\begin{document}

\def\levelset{{\cal X}}
\def\fxx{{\bf x}}
\def\fxy{{\bf y}}
\def\real{I\!\!R}

\begin{center}
{\Large Equivalence of Descriptions of Gravity in Both Curved and Flat Space-time }\\
\vskip 0.2in
Mei Xiaochun\\
\vskip 0.2in
( Institute of Theoretical Physics in Fuzhou, No.303, Building 2, Yinghu Garden,\\
Xihong Road, Fuzhou, 350025, P.R.Chian, E-mail: fzbgk@pub3.fz.fj.cn )\\
\end{center}

\begin{abstract}
It is proved in the manuscript that as long as the proper coordinate transformation is introduced, the equations of geodetic lines described in curved space-time can be transformed into the dynamic equations in flat space-time. That is to say, the Einstain theory of gravity and other gravitational theories based on the curved space-time can be identically transformed into flat space-time to describe. As an example, the Schwarzschild solution of the spherical symmetry gravitational field is transformed into flat space-time to study. The results show that there exists no any singularity in the all processes and the whole space-time including at the point r=0. So it seems more rational to discuss the problems of gravitation in flat space-time.
\par
PACS number 0400\\
\end{abstract}
\begin{center}
{\large Introduction}\\
\end{center}
\par
    The general theory of relativity based on the curved space-time has got great success and becomes main current theory now. However, there still exist some foundational problems in it just as the definition of gravitational field's energy, the quantization of gravitation and the problem of singularity and so on. So it is always an attractive idea to re-establish gravitational theory in flat space-time. Since the 1940's, many theories based on flat space-time were put forward. Though all of those theories are coincident with the gravitational theory of Einstain under the conditions of weak fields, it can not be proved that they are better than the theory of Einstain by experiments at present. So according to the current viewpoint, the space-time of gravitation field should be non-Euclidean one. The flat space-time is always regarded as the boundary condition where the gravitational field is far away.
\par
    In the paper, the author does not try to establish an independent theory in flat space-time. But it can be proved that as long as the proper coordinate transformations are introduced and, the equations of geodetic lines in curved space-time can be transformed into the dynamic equations in flat space-time. It means that the Einstein's theory of gravitation and other theories based on the curved space-time can also be identically transformed into the flat space-time to describe. Then, the method is used to discuss the Schwarzschild solution of the spherical symmetry gravitational field. The results show that there exists no any singularity again in the all processes and the whole space-time including the point . So it seams more rational to discuss gravitational problems in the flat space-time.
\par
The paper includes three chapters. The first chapter discusses how to transform the Schwarzschild solution of Einstein's theory in the spherical symmetry gravitation field into the dynamic equations in flat space-time. The second chapter provides a general proof to transform all gravitational theories based on curved space-time into the theories based on flat space-time. The third chapter discusses some problems of foundational concepts, for example, weather space-time is curved or not when exit gravitational fields exist.\\
\begin{center}
{\large 1. The Transformation of the Schwarzschild Solution}\\
\end{center}
\par
According to the general theory of relativity, the Schwarzschild metric of the spherical symmetry gravitation field is
\begin{equation}
ds^2=c^2(1-{\alpha\over{r}})dt^2-(1-{\alpha\over{r}})^{-1}{dr^2}-r^2(\sin^2\theta{d}\varphi^2+d\theta^2)
\end{equation}
In the formula, we take $\alpha=2GM/c^2$ . Let $\theta=\pi/2$ and put Eq.(1) into the geodetic line equation, according to the familiar results in the general theory of relativity, we have the integrals
\begin{equation}
c(1-{\alpha\over{r}}){dt\over{ds}}=\varepsilon
\end{equation}
\begin{equation}
r^2{d\varphi\over{ds}}={L\over{c}}
\end{equation}
Here $\varepsilon$ and $L$ are constants. From above two formulas, the linear element $ds$ can be eliminated and we can get 
\begin{equation}
r^2(1-{\alpha\over{r}})^{-1}{d\varphi\over{dt}}={L\over{\varepsilon}}
\end{equation}
Defining
\begin{equation}
d\tau=(1-{\alpha\over{r}})dt
\end{equation}
and regarding $\tau$ as the eigen time , $t$ as the coordinate time and taking  $\varepsilon=1$,we can write Eq.(2)as
\begin{equation}
cd\tau=ds
\end{equation}
Eq.(4) becomes
\begin{equation}
r^2{d\varphi\over{d\tau}}=L
\end{equation}
Here $L$ is the angel momentum of unit mass. Eq.(7) is just the conservation formula of angel momentum. 
\par
    Let's first discuss the motions of particles with static masses. By using Eq.(6), Eq.(1) can be written as 
\begin{equation}
(1-{\alpha\over{r}})({dt\over{d\tau}})^2-{1\over{c^2}}(1-{\alpha\over{r}})^{-1}({dr\over{d\tau}})^2-{r^2\over{c^2}}({d\varphi\over{d\tau}})^2=1
\end{equation}
By using Eq.(5) and (7), we get
\begin{equation}
({dr\over{d\tau}})^2={c^2\alpha\over{r}}(1-{L^2\over{\alpha{c^2}r}}+{L^2\over{c^2r^2}})
\end{equation}
Taking the differential about $d\tau$ in the formula above, we get
\begin{equation}
{d^2r\over{d\tau^2}}-{L^2\over{r^3}}=-{c^2\alpha\over{2r^2}}(1+{3L^2\over{c^2r^2}})
\end{equation}
It should be noted that each quantity in Eq.(10) is defined in caved space-time. In order to express the equation in flat space-time, the further transformation is needed. Let $r_0$, $\varphi_0$ and $t_0$ represent the space-time coordinates in flat space-time, because of the invariability of the 4-diamention interval $ds^2$ , we have
\begin{equation}
ds^2=c^2dt^2_0-dr^2_0-r^2_0{d}\varphi^2_0=c^2(1-{\alpha\over{r}})dt^2-(1-{\alpha\over{r}})^{-1}dr^2-r^2d{\varphi^2}
\end{equation}
Let $r_0=r$,$\varphi_0=\varphi$ £¬ we get the transformation relation between $t_0$ and $t$
\begin{equation}
c^2dt^2_0=c^2(1-{\alpha\over{r}})dt^2+[1-(1-{\alpha\over{r}})^{-1}]dr^2
\end{equation}
Considering Eq.(5) and (9), we get
\begin{equation}
dr=c(1-{\alpha\over{r}})\sqrt{{\alpha\over{r}}(1-{L^2\over{\alpha{c^2}r}}+{L^2\over{c^2r^2}})}dt
\end{equation}
Put it into Eq.£¨12£©, we have
\begin{equation}
dt_0=\sqrt{(1-{\alpha\over{r}})[1-{\alpha^2\over{r^2}}(1-{L^2\over{\alpha{c^2}r}}+{L^2\over{c^2r^2}})]}dt
\end{equation}
Comparing it with Eq.(5), we get
\begin{equation}
d\tau=(1-{\alpha\over{r}})^{1\over{2}}[1-{\alpha^2\over{r^2}}(1-{L^2\over{\alpha{c^2}r}}+{L^2\over{c^2r^2}})]^{-{1\over{2}}}dt_0
\end{equation}
Combining Eq.(7)with (10) and let $r_0\rightarrow{r}$£¬the results of the Einstein's theory can be expressed in the similar formula of the Newtonian gravitation in flat space-time
\begin{equation}
{d^2\vec{r}\over{d\tau^2}}=-GM(1+{{3L^2}\over{c^2r^2}}){\vec{r}\over{r^3}}
\end{equation}
Let $u=1/r$ and by considering Eq.(7), Eq.(16) can be transformed into
\begin{equation}
{{d^2u}\over{d\varphi^2}}+u={c^2\alpha\over{2L^2}}(1+{{3L^2}\over{c^2}}u^2)
\end{equation}
The formula can describe the perihelion precession of the Mercury. 
\par
    On the other hand, We have used the eigen time $\tau$ in Eq.(16). It can be proved that the effect of special relativity has been considered in the formula. The square of a particle's speed in the center gravitational field is
\begin{equation}
V^2=V^2_r+V^2_{\varphi}
\end{equation}
From Eq.(7),(9) and (15), we have
$$V^2_r=({dr\over{dt_0}})^2=({dr\over{d\tau}}{d\tau\over{dt_0}})^2={{c^2\alpha}\over{r}}(1-{\alpha\over{r}})(1-{L^2\over{\alpha{c^2}r}}+{L^2\over{c^2r^2}})[1-{\alpha^2\over{r^2}}(1-{L^2\over{\alpha{c^2}r}}+{L^2\over{c^2r^2}})]^{-1}$$
\begin{equation}
V^2_{\varphi}=(r{{d\varphi}\over{dt_0}})^2=(r{{d\varphi}\over{d\tau}}{{d\tau}\over{dt_0}})^2={L^2\over{r^2}}(1-{\alpha\over{r}})[1-{\alpha^2\over{r^2}}(1-{L^2\over{\alpha{c^2}r}}+{L^2\over{cr^2}})]^{-1}
\end{equation}
Therefore, we have
$$V^2=V^2_r+V^2_{\varphi}={c^2\alpha\over{r}}(1-{\alpha\over{r}})(1+{L^2\over{c^2r^2}})[1-{\alpha^2\over{r^2}}(1-{L^2\over{\alpha^2{c^2}r}}+{L^2\over{c^2r^2}})]^{-1}$$
\begin{equation}
1-{V^2\over{c^2}}=(1-{\alpha\over{r}})[1-{\alpha^2\over{r^2}}(1-{L^2\over{\alpha^2{c^2}r}}+{L^2\over{c^2r^2}})]^{-1}
\end{equation}
Comparing Eq.(20) with Eq.(15), we obtain
\begin{equation}
d\tau=\sqrt{1-{v^2\over{c^2}}}dt_0
\end{equation}
It is completely the same as the formula of time delay in the special theory of relativity. Therefore, Eq.(16) can be written as
\begin{equation}
{d\vec{p}\over{dt}}=-GMm(1+{{3L^2}\over{c^2r^2}})\sqrt{1-{V^2\over{c^2}}}{\vec{r}\over{r^3}}
\end{equation}
Here $m$ is the mass of particle, and $d\tau$ is determined by Eq.(15). Because of $\vec{L}\rightarrow\vec{V}\times\vec{r}$, it can be seen that there exist the extra two items relative to $V^2/c^2$ comparing with The Newtonian theory.
\par
The problem of energy conservation is discussed below. For simplification, we only discuss the situation with $L=0$ . In this case, the particle moves along the radium direction. By using Eq.(20), Eq.(22) becomes
\begin{equation}
{d\vec{p}\over{dt}}=-{GMm\over{\sqrt{1+{\alpha/r}}}}{\vec{r}\over{r^3}}
\end{equation}
By producing $d\vec{r}$ on the two sides of Eq.(23) and taking the integral, we have
\begin{equation}
\int{d\vec{p}\over{dt}}\cdot{d\vec{r}}=-\int{GMm\over{\sqrt{1+\alpha/r}}}\cdot{d\vec{r}}
\end{equation}
The left side of Eq.(24) can be written as
\begin{equation}
\int{d\vec{p}\over{dt}}\cdot{d\vec{r}}=\int{d\vec{p}\over{dt}}\cdot{d\vec{r}\over{dt}}dt=\int\vec{V}\cdot{d\vec{p}}=\int\vec{V}\cdot{d}{m\vec{V}\over{\sqrt{1-V^2/c^2}}}={mV^2\over{\sqrt{1-V^2/c^2}}}+mc^2{\sqrt{1-{V^2\over{c^2}}}}+E_1
\end{equation}
Here $E_1$ is a constant. So Eq.(24)can be written as
\begin{equation}
{mV^2\over{\sqrt{1-V^2/c^2}}}+mc^2{\sqrt{1-{V^2\over{c^2}}}}=mc^2{\sqrt{1+{\alpha\over{r}}}}+E
\end{equation}
Here $E$ is a constant. Supposes $V\rightarrow{0}$ when $r\rightarrow\infty$,we have $E=0$. Eq.(26) can be written as
\begin{equation}
{mV^2\over{\sqrt{1-V^2/c^2}}}+mc^2({\sqrt{1-{V^2\over{c^2}}}}-1)+mc^2(1-\sqrt{1+{\alpha\over{r}}})=0
\end{equation}
Let $T$ represents the kinetic energy of the particle, $U$ represents the potential energy of the particle. We define
\begin{equation}
T={mV^2\over{\sqrt{1-V^2/c^2}}}+mc^2(\sqrt{1-{V^2\over{c^2}}}-1)
\end{equation}
\begin{equation}
U=mc^2(1-\sqrt{1+{\alpha\over{r}}})
\end{equation}
Eq.(27)is just the formula of energy conservation $T+U=E=0$ . When $\alpha/r<<1$ , $V<<c$ £¬from Eq.(27£© we get the result of the Newtonian theory.
\begin{equation}
{mV^2\over{2}}-{GMm\over{r}}=0
\end{equation}
\par
    The motion equation of photon in the center gravitational field is discussed as follows. For photons,$ds=0$ £¬so $ds$ can not be used as the parameter of the equation of geodetic line. In this case, we take $d\tau$ to replace $ds$ and get the same results by solving the equations of gravitational field
\begin{equation}
(1-{\alpha\over{r}}){dt\over{d\tau}}=\varepsilon
\end{equation}
\begin{equation}
r^2{d\varphi\over{d\tau}}=L
\end{equation}
Let $\varepsilon=1$£¬we have
\begin{equation}
d\tau=(1-{\alpha\over{r}})dt
\end{equation}
Because
\begin{equation}
ds^2=c^2(1-{\alpha\over{r}})dt^2-(1-{\alpha\over{r}})^{-1}dr^2-r^2d\varphi^2=0
\end{equation}
We get
\begin{equation}
c^2(1-{\alpha\over{r}})({dt\over{d\tau}})^2-(1-{\alpha\over{r}})^{-1}({dr\over{d\tau}})^2-(r{d\varphi\over{d\tau}})^2=0
\end{equation}
From the formula above we have
\begin{equation}
({dr\over{d\tau}})^2=c^2[1-(1-{\alpha\over{r}}){L^2\over{c^2r^2}}]
\end{equation}
Taking the differential about $d\tau$, we get
\begin{equation}
{d^2r\over{d\tau^2}}-{L^2\over{r^3}}=-{3\alpha{L}^2\over{2r^4}}
\end{equation}
By using Eq.(33) and (36), we get
\begin{equation}
({dr\over{dt}})^2=c^2(1-{\alpha\over{r}})^2[1-(1-{\alpha\over{r}})
{L^2\over{c^2r^2}}]
\end{equation}
Suppose the speed of photon in the gravitational field is $V$, from Eq.£¨32£©£¬£¨33£©and£¨38£©we have
\begin{equation}
V=\sqrt{({dr\over{dt}})^2+(r{d\varphi\over{dt}})^2}=c(1-{\alpha\over{r}})\sqrt{1+{\alpha{L^2}\over{c^2r^3}}}
\end{equation}
It is obvious that $V\neq$ constant, so the speed of light would change with $r$ in gravitational field. Then let's discuss how to transform the results into the flat reference system. For light's motion, if we write the metric in the flat reference system as
\begin{equation}
ds^2=c^2dt^2_0-dr^2_0-dr^2_0-r^2_0{d}\varphi^2_0=0
\end{equation}
the result shows that light move in a uniform speed $c$ in the gravitational field. However, this is improper for it contradicts Eq.(39). Suppose light's speed is $V_0$ in the flat space-time, the metric should be written as
\begin{equation}
ds^2=u^2_0{d}t^2_0-dr^2_0-r^2_0{d}\varphi^2_0=0
\end{equation}
From Eq.(34) and (41)we have
\begin{equation}
V^2_0{d}t^2_0-dr^2_0-r^2_0{d}\varphi^2_0=c^2(1-{\alpha\over{r}})dt^2-(1-{\alpha\over{r}})^{-1}dr^2-r^2d\varphi^2
\end{equation}
Let $r_0=r$,$\varphi_0=\varphi$ we get from Eq.(42)
\begin{equation}
V^2_0{d}t^2_0=c^2(1-{\alpha\over{r}})dt^2+[1-(1-{\alpha\over{r}})^{-1}]dr^2
\end{equation}
By using Eq.(38), we have
\begin{equation}
V^2_0{d}t^2_0=c^2(1-{\alpha\over{r}})^2(1+{\alpha{L}^2\over{c^2r^3}})dt^2
\end{equation}
There exists one degree of freedom to choose the relations between $t_0$ and $t$ here. If taking $t_0=t$, we get
\begin{equation}
V_0=c(1-{\alpha\over{r}}){\sqrt{1+{\alpha{L^2}\over{c^2r^3}}}}
\end{equation}
Comparing Eq.(45) with Eq.(39) we obtain $V_0=V$, that is to say, the speeds of lights are the same in the both curved and flat space-times. Therefore, by the relation $\alpha=dV_0/dt_0=dV/dt$, the accelerations and forces are the same, so that the forms of motion equations are also the same. So by connecting Eq.(32) with (37), we can directly write the motion equation of photons in the vector's form in flat space-time as
\begin{equation}
{d^2\vec{r}\over{d\tau^2}}=-{3\alpha{L^2}\vec{r}\over{2r^5}}
\end{equation}
Let $u=1/r$ , the formula can be transformed into
\begin{equation}
{d^2u\over{d\varphi^2}}+u={3\alpha\over{2}}u^2
\end{equation}
The formula can be used to describe the deviation of light in the solar gravitational field. As for the time delay experiments of radar waves in the solar gravitational field, by considering $t_0=t$ in Eq.(38) under the condition of week field, we can get
\begin{equation}
cdt_0=(1+{\alpha\over{r}})(1-{L^2\over{r^2}})^{-{1\over{2}}}(1-{\alpha{L^2}\over{c^2r^3}})dr
\end{equation}
Suppose radar waves just swept over the surface of the sun with the radium $r_0$ and the speed of radar waves is the speed of light in vacuum, we have $L=cr_0$ in the light of angular conservation. The integral of Eq.(48) is
\begin{equation}
ct_0=\sqrt{r^2-r^2_0}+\alpha{l}n{\sqrt{r^2-r^2_0}+r\over{r_0}}-\alpha{\sqrt{r^2-r^2_0}\over{2r}}
\end{equation}
The same result can be reached from the formula $^{(2)}$ .
\par
However, it can be seen from Eq.(38) that the photon would move in the speed over light' speed in vacuum when $L=0$ and $r<\alpha/2$. This is unacceptable (The problem will be discussed again later.). So we re-define the transform relations between $t_0$ , $t$ and $\tau$ as£º
\begin{equation}
dt_0=(1-{\alpha^2\over{r^2}})dt=(1+{\alpha\over{r}})d\tau
\end{equation}
Put it into Eq.(36), we have
\begin{equation}
{dr\over{dt_0}}=c(1+{\alpha\over{r}})^{-1}\sqrt{1-(1-{\alpha\over{r}}){L^2\over{c^2r^2}}}
\end{equation}
From the formula, Eq.(49) can also be reached under the condition of weak field. It is obvious that Eq.(50) is the simplest form to obtain the formula (49).from (36). In this way, we have
\begin{equation}
V_0=c\sqrt{1+{\alpha{L^2}\over{c^2r^3}}}(1+{\alpha\over{r}})^{-1}
\end{equation}
Because $\vec{L}=\vec{V_0}\times\vec{r}$ ,when $r\rightarrow{0}$ , $V_0\rightarrow{0}$ , there is no the motion of over light's speed again according to Eq.(52).
\par
It should be noted that Eq.£¨46£©is not the dynamic equation of the photon in the gravitational field. Because when $L=0$ , $d^2\vec{r}/d\tau^2=0$ , it seems that the photon is not acted by force. However, the photon has acceleration in the gravitational field. So there should be a force acting on the photon. Therefore, Eq.£¨46£©can only be regarded as the equation of kinematics from which the velocity and acceleration of the phone can be obtained. But it can not be regarded as the equation of dynamics of the photon from which the force can be obtained. 
\par
    Now let's discuss how to obtain the dynamic equation of the photon based on Eq.(46). Let photon's speed $V_0\rightarrow{V}$ $($as well as $r_0\rightarrow{r}$ , $t_0\rightarrow{t}$ $)$ , when $r\rightarrow\infty$, $V\rightarrow{c}$ . When $r<\infty$ , $V<c$ , showing that the photon is acted by repulsion and does retarded motion. In order to obtain the dynamic equation of photon in the gravitational field, we suppose to have an imaginary particle with speed $\vec{V'}$ and
\begin{equation}
\vec{V}'=\vec{c}-\vec{V}
\end{equation}
Here $V$ is the speed of photon in the gravitational field, $c$ is the speed of photon in vacuum. The directions of $V$ and $c$ are supposed always the same. When photon's initial speed $V=c$ , the speed of imaginary particle is $V'=0$ . When the photon falls down in the gravitational field with $V<c$ , we have $V'>0$ . When $V=0$ , we have $V'=c$ . So it is obvious that the imaginary particle does acceleration motion in the gravitational field similar to the general particles with static masses. Therefore, the force acted on the imaginary particle in the gravitational field of spherical symmetry can be supposed to be
\begin{equation}
\vec{F}'=-GMm(1+{3L^2\over{c^2r^2}})\sqrt{1-{V'^2\over{c^2}}}{\vec{r}\over{r^3}}
\end{equation}
Here $m$ is the static mass of the imaginary particle. The dynamic equation of imaginary particle is
\begin{equation}
{d\vec{p}'\over{dt}}=-GMm(1+{3L^2\over{c^2r^2}})\sqrt{1-{(\vec{c}-\vec{V})^2\over{c^2}}}{\vec{r}\over{r^3}}
\end{equation}
On the other hand, similar to the particles with static masses, the relativity momentum of a photon in the gravitational field can be defined as
\begin{equation}
\vec{p}={m\vec{V}\over{R}}
\end{equation}
Here $m$ is so-called photon's static mass and $R=R(r,\theta)$ is the function remained to be decided. Then, we define the imaginary particle's momentum $\vec{p}'$ as 
\begin{equation}
\vec{p}'=\vec{p}_c-\vec{p}=m\vec{c}-{mV\over{R}}
\end{equation}
Here $\vec{p}_c$ is photon's momentum in vacuum. Put Eq.(57) into Eq.(55), we get the dynamic equation of the photon in the gravitational field of spherical symmetry
\begin{equation}
{d\vec{p}\over{dt}}=GMm(1+{3L^2\over{c^2r^2}})\sqrt{{{2cV-V^2}\over{c^2}}}{\vec{r}\over{r^3}}=\vec{F}
\end{equation}
\par
However, it can be seen that it is actually unnecessary for us to introduce imaginary particle. In fact, we can directly suppose that the dynamic equation of photon in the central gravitational field is just Eq.(58), so long as from it we can reach the identical results comparing with the Einstein's theory. The concrete form of function $R$ is discussed as follows. We have from Eq.(56)
\begin{equation}
{d\vec{p}\over{dt}}={d\over{dt}}{m\over{R}}{d\vec{r}\over{dt}}={m\over{R}}{d^2\vec{r}\over{dt^2}}+m\vec{V}{d\over{dt}}{1\over{R}}
\end{equation}
Put it into Eq.(58), we get photon's acceleration in the gravitational field
\begin{equation}
{d^2\vec{r}\over{dt^2}}=R{\vec{F}\over{m}}-R\vec{V}{d\over{dt}}{1\over{R}}
\end{equation}
On the other hand, from Eq.(46) and (50), we can get the result of the Einstein's theory in the flat space-time
\begin{equation}
{d^2\vec{r}\over{dt^2}}=-{3\alpha{L^2}\vec{r}\over{2r^5}}(1+{\alpha\over{r}})^{-2}+{\alpha{V_r}\vec{V}\over{r^2}}(1+{\alpha\over{r}})^{-1}
\end{equation}
Comparing Eq.(60) with (61), we get 
\begin{equation}
R({\vec{F}\over{m}}-\vec{V}{d\over{dt}}{1\over{R}})=-{3\alpha{L^2}\vec{r}\over{2^2r^5}}(1+{\alpha\over{r}})^{-2}+{\alpha{V_r}\vec{V}\over{r^2}}(1+{\alpha\over{r}})^{-1}
\end{equation}
Decomposing the formula in the both directions of $\vec{e}_{r}$ and $\vec{e}_{\varphi}$ ,we have
\begin{equation}
R({F\over{m}}-V_r{d\over{dt}}{1\over{R}})=-{3\alpha{L^2}\over{2^2r^4}}(1+{\alpha\over{r}})^{-2}+{\alpha{V^2_r}\over{r^2}}(1+{\alpha\over{r}})^{-1}
\end{equation}
\begin{equation}
RV_{\varphi}{d\over{dt}}{1\over{R}}={\alpha{V_r}V_{\varphi}\over{r^2}}(1+{\alpha\over{r}})^{-1}
\end{equation}
Putting Eq.(64)into(63) and using Eq.(52), we get
\begin{equation}
R={3L^2\over{c^2r^2}}(1+{\alpha\over{r}}+{3L^2\over{r^2}}+{3\alpha{L^2}\over{C^2r^3}})^{-1}(1+{2\alpha\over{r}}+{\alpha{L^2}\over{c^2r^3}}+
{2\alpha^2{L^2}\over{c^2r^4}})^{-{1\over{2}}}
\end{equation}
When $L=0$ we have $V=V_r$ ,$V_{\varphi}=0$ . In this case, Eq.(64) does not exist. But from Eq.(63)directly, we have
\begin{equation}
{dR\over{dr}}=-{\alpha\over{2r^2}}\sqrt{1+{2\alpha\over{r}}}(1+{\alpha\over{r}})R^2+{\alpha\over{r^2}}R
\end{equation}
This is the quasi-one order Bernoulii equation£¬the solution is
\begin{equation}
R=[-e^x\int\sqrt{1+2x}(1+x)e^{-x}dx+C]^{-1}
\end{equation}
Here $x=\alpha/r$ When $x\rightarrow$ , $R=1$£¬the integral constant C can be determined.
\par
After the function $R$ is determined by the method above, Eq.(58) can be regarded as the dynamic equation of photon in the gravitational field of spherical symmetry in flat space-time. It coincides with the results of the Einstein's theory and can be used to explain the deviation of light as well as the time delay experiments of radar waves in the solar gravitational field. Besides, Eq.(58) can also be used to explain the gravitational red shift of spectral line. Let's first establish the energy conservation equation of photon in the gravitational field. Similar to Eq.(24), we multiply $d\vec{r}$ on the two sides of Eq.(58) and take the integral
\begin{equation}
\int{d\vec{p}\over{dt}}\cdot{d}\vec{r}=\int{G}Mm(1+{3L^2\over{c^2r^2}})\sqrt{{{2cV-V^2}\over{c^2}}}{\vec{r}\over{r^3}}\cdot{d}\vec{r}
\end{equation}
We only considering the situation with $L=0$, by using Eq.(52), the two sides of Eq.(68) can be written as
\begin{equation}
{mc^2\over{R}}(1+{\alpha\over{r}})^{-2}-\int{mc^2\alpha\over{R^2r}}(1+{\alpha\over{r}})^{-3}dr=\int{mc^2\alpha\over{2r^2}}(1+{\alpha\over{r}})^{-1}\sqrt{1+{2\alpha\over{r}}}dr
\end{equation}
The integral of the right side is
\begin{equation}
\int{mc^2\alpha\over{2r^2}}(1+{\alpha\over{r}})^{-1}\sqrt{1+{2\alpha\over{r}}}dr=-mc^2(\sqrt{1+{2\alpha\over{r}}}-\alpha{r}ctg\sqrt{1+{2\alpha\over{g}}})+C_2
\end{equation}
When $r\rightarrow\infty$ , we have
\begin{equation}
-mc^2(\sqrt{1+{2\alpha\over{r}}}-\alpha{r}ctg\sqrt{1+{2\alpha\over{g}}})_{r\rightarrow\infty}=-mc^2(1-{\pi\over{4}})
\end{equation}
So when $L=0$ we can define the potential energy of photon in the central gravitational field as 
\begin{equation}
U(r)=mc^2(\sqrt{1+{2\alpha\over{r}}}-\alpha{r}ctg\sqrt{1+{2\alpha\over{g}}}-1+{\pi\over{4}})
\end{equation}
When $r\rightarrow\infty$ we have $U(r)\rightarrow{0}$. Let
\begin{equation}
{mc^2\over{R}}(1+{\alpha\over{r}})^{-2}-\int{mc^2\alpha\over{Rr^2}}(1+{\alpha\over{r}})^{-3}dr=K(r)+C_1
\end{equation}
Eq.(69) can be written as
\begin{equation}
K(r)+mc^2(1-{\pi\over{4}})+mc^2(\sqrt{1+{2\alpha\over{r}}}-\alpha{r}ctg\sqrt{1+{2\alpha\over{r}}}-1+{\pi\over{4}})=C
\end{equation}
Here $C$ is a constant. If we choose another proper constant $b$ and let $V_0$ represent the light's frequency in vacuum, and define the total energy of photon as $E=C+b=hv_0=mc^2$ , the kinetic energy of photon can be defined as
\begin{equation}
T(r)=K(r)+mc^2(1-{\pi\over{4}})+b
\end{equation}
In this way, when $L=0$ , the formula of energy conservation of photon in the gravitational field of spherical symmetry can be written as
\begin{equation}
T(r)+U(r)=E=hv_0
\end{equation}
On the other hand, in the general theory of relativity, the red shift of spectral line is considered caused by time delay of gravitational field. In the gravitational theory of curved space-time, the photon is actually considered to be free one and moves along the curved geodetic line without potential energy or no force acted on it. Therefore, the total energy of a photon is equal to its kinetic energy in the gravitational field. But if the formula $E=hv$ is considered tenable in any point in the gravitational field, because  is a variable, is also a variable, that is to say, that the energy of photon in the gravitational field is not conservative. This is just the price we have to pay in the gravitational theory based on the curved space-time to explain the red shift of spectral line, thought now people seem to avoid this problem. However, this is unacceptable when we describer gravitational force in flat space-time. In order to keep energy conservation for photon in the description of flat space-time, the rational way is to suppose that the frequency of photon is only relative to its kinetic energy, and has nothing to do with its potential energy with $T=hv$ . Here $v$ is the frequency of photon in the point  in the field. So the formula of energy conservation of photon in the gravitational field is $hv+U(r)=hv_0=mc^2$ . In this way, the formula of red shift becomes
\begin{equation}
{\bigtriangleup{v}\over{v_0}}={{v-v_0}\over{v_0}}=-(\sqrt{1+{2\alpha\over{r}}}-\alpha{r}ctg\sqrt{1+{2\alpha\over{r}}}-1+{\pi\over{4}})
\end{equation}
Under the condition of weak field with $\alpha/r<<1$ £¬by the developing of the Taylor series, we have
\begin{equation}
{\bigtriangleup{v}\over{v_0}}={{v-v_0}\over{v_0}}=-({\alpha\over{r}}-{\alpha\over{2r}})=-{GM\over{r}}
\end{equation}
Under the same condition, the result of the general theory of relativity is
\begin{equation}
{\bigtriangleup{v}\over{v_0}}={{v-v_0}\over{v_0}}=\sqrt{1-{\alpha\over{r}}}-1=-{GM\over{r}}
\end{equation}
The both are the same. But in the strong field, they are not the same, especially when $\alpha>r$, Eq.(79) becomes meaningless, but Eq.(77)is still meaningful. In fact Eq.(77) can be used to explain the big red shifts of the quasi-stellar objects. Taking $\alpha/r=7.5$, we have the red shift value $Z=-\bigtriangleup{v}/v_0=2.47$ . Suppose the mass of the quasi-stellar object is $10^{40}Kg$, it can be calculated that the radium of he quasi-stellar object is $r=1.98\times{10}^{12}m$ and its mean density is $\rho=3.08\times{10}^2Kg/m^3$ . It is only $0.22$ times comparing with the mean density of the sun with $\rho=1.40\times{10}^3Kg/m^3$ . Taking $\alpha/r=12$ £¬we have $Z=3.42$ , $r=1.24\times{10}^{12}m$ , $\rho=1.26\times{10}^2Kg/m^3$ . The even density is similar to the sun's density. According to the general theory of relativity, if $\alpha/r>1$ £¬ $r$ is in the inside of black hole. But according to Eq.(77), only when $r\rightarrow{0}$ , we have infinite red shift with $Z\rightarrow\infty$ . For the common stars, $r\neq{0}$ £¬so according to Eq.(77), the black holes do not exist actually.
\par
What is mainly shown above is that the results are the same under the condition of weak field when the gravitational theory is described in the both flat and curved space-time. Besides the red shift of spectral line, the following discussion will further show their differences in the strong fields owing to the fact that the coordinate times are different in the both situations. We also only discuss particle's motions along the radial direction of the spherical symmetry gravitational field with $L=0$ . For a particle with static mass, when $L=0$ , the speed is from Eq.(9) and (5)
\begin{equation}
V={dr\over{dt}}=\pm{c}\sqrt{{\alpha\over{r}}}(1-{\alpha\over{r}})
\end{equation}
In the formula it has been supposed that $V=0$ when $r\rightarrow\infty$ . Definite the direction of velocity is positive along the radius vector's positive direction. The direction of particle's velocity is negative when the particle falls down the gravitational field. The acceleration is 
\begin{equation}
\alpha={dV\over{dt}}=-{1\over{2}}{c^2\alpha\over{r^2}}(1-{\alpha\over{r}})(1-{3\alpha\over{r}})
\end{equation}
Take the integral of Eq.£¨80£©and suppose $r=r_0$ when $t=0$ , we have
\begin{equation}
ct=\pm{1\over{\sqrt{\alpha}}}[{2\over{3}}(r^{3\over{2}}-r^{3\over{2}}_0)+2\alpha(\sqrt{r}-\sqrt{r_0})+{\alpha}^{3\over{2}}ln{{({\sqrt{r}-\sqrt{\alpha}})({\sqrt{r_0}}+{\sqrt{\alpha}})}\over{({\sqrt{r}+\sqrt{\alpha}})({\sqrt{r_0}-\sqrt{\alpha}})}}]
\end{equation}
                   (82)
Because $t>0$ £¬the negative sign is taken when a particle falls down along the direction of radius vector. Conversely, it takes positive sign.
\par
Let's first discuss the particle's motion in the area $r\geq\alpha$ . When , $r>a$, $V$ is a negative number. When $r=\alpha$ , $V=0$ £¬the particle arrives at the Schwarzschild event horizon. It can be known from Eq.(81) that when $r=3\alpha$ and $r=\alpha$ , acceleration  becomes zero. When $r>3\alpha$ ,$\alpha<0$ £¬the particle is acted by gravitation and is accelerated downward. When $\alpha<r<3\alpha$ , $\alpha>0$ , the particle is acted by repulsive force and is retarded £¨Gravitation becomes repulsion, it seems unimaginable.£©.When $r=\alpha$ , the particle is not acted by force and at rest on the surface of event horizon. It can be known from Eq.(82) that when $r=\alpha$ , $t=\infty$ , that is to say, it takes the particle an infinite time to reach the event horizon.
\par
Then let's discuss the particle's motion beneath the event horizon $r<\alpha$ . It can be known from Eq.(82) that the time has no definition inside the event horizon because the logarithm of a negative number has no definition(At present, some peasants think that it means space and time to be exchanged each other in the black holes. This is absurd. How one dimension's time can be transformed into three dimension's space?). So speaking strictly, it is meaningless to talk about particle's speed, acceleration and motion in the area $r<\alpha$ . If this problem is neglected temporary, we can write Eq.(80)as
\begin{equation}
V=\pm{c}\sqrt{{\alpha\over{r}}}(1-{\alpha\over{r}})+V_0
\end{equation}
When $r=\alpha$ , $V=0$ £¬so $V_0=0$ . Therefore, the velocity and acceleration can still be expressed by Eq.(80) and (81). It is obvious that we have $\alpha<0$ inside the event horizon, meaning that the particles only acted by gravitation. Suppose a particle at the point $r$ has a velocity upward, it would be retarded until $a=V=0$ when it arrives at the event horizon and stays there at last. If a particle has a velocity downward, is would be accelerated and reach light's speed at a certain place. After that, the particle would move in the speed over light's speed and reaches an infinite speed at the point $r=0$. The results is the same as that analyzed in the current theory using the method of light cone, except that the particle's speed would be over light's speed.
\par
As for a photon's motion, when $L=0$, from Eq.(38) we have
\begin{equation}
V={dr\over{dt}}=\pm{c}(1-{\alpha\over{r}})
\end{equation}
\begin{equation}
\alpha={dV\over{dt}}={c^2\alpha\over{r^2}}(1-{\alpha\over{r}})
\end{equation}
Let $r=r_0$ when $t=0$ , we have integral from Eq.(84)
\begin{equation}
ct=\pm(r-r_0+\alpha{l}n{{r-\alpha}\over{r_0-\alpha}})
\end{equation}
When a photon falls down in the area $r>\alpha$ , because $\alpha>0$ £¬they are retarded by repulsion. When the photon arrives at the event horizon $r=\alpha$ , the speeds and accelerations are equal to zero and the infinite time is need. In the area $r<\alpha$ , time has no definition for the same reason. Despite of this problem, we have $\alpha<0$ in the area $r<\alpha$. Suppose a photon has a velocity upwards inside the event horizon, it would be retarded by gravitation and has $\alpha=V=0$ when it arrives at the surface of event horizon. If the photon has a velocity downwards, it would be accelerated. When it arrives at the point $r=\alpha/2$ , its speed would reach light's speed in vacuum again. After that time, the photon would move in the speed over light's speed in vacuum. When the photon arrives at the point $r=0$ , its speed becomes infinite $^{(3)}$.
\par
     It is obvious that there exist some things irrational, especially particles would move in the speeds over light's speed in vacuum. In fact in the current theory of black holes, the motions with the speeds over light's speed can not be avoided during the processes in which material collapses toward the center singularities of gravitational fields. At present, those problems are attributed to the improper selections of coordinates. In order to eliminate those defects, people now transform the problems into other coordinate systems to discuss, for example, the Eddington and the Kruskal coordinate system. In the new coordinate system, though the singularities on the surfaces of event horizons can be eliminated, they can not yet be eliminated at the point $r=0$. Hawking even proved that it was impossible to eliminate all singularities in the general theory of relativity $^{(3)}$.
\par
Now let's discuss the problems in flat space-time. When $t_0=0$ let $r=r_0$ . According to Eq.(19), when a particle falls free down the gravitational field, we have
\begin{equation}
V=-c\sqrt{{\alpha\over{r}}}(1+{\alpha\over{r}})^{-{1\over{2}}}
\end{equation}
\begin{equation}
\alpha=-{1\over{2}}{c^2\alpha\over{r^2}}(1+{\alpha\over{r}})^{-2}
\end{equation}
\begin{equation}
ct_0={2\over{3\sqrt{\alpha}}}[(r_0+\alpha)^{3\over{2}}-(r+\alpha)^{3\over{2}}]
\end{equation}
It is obvious that every thing is normal in the area $r>\alpha$ . The particle is monotonously accelerated by gravitation. There is no any singularity in the whole space-time and in all physical quantities. When the particle arrives at the point $r=0$ , we have
\begin{equation}
V=-\lim_{x\rightarrow\infty}{c\sqrt{x}\over{\sqrt{1+x}}}\rightarrow{-c}~~~~~~(x={\alpha\over{r}})
\end{equation}
\begin{equation}
\alpha=-\lim_{x\rightarrow\infty}{c^2x^2\over{2\alpha(x+1)^2}}\rightarrow{-}{c^2\over{2\alpha}}
\end{equation}
It can be seen that the particle's speed tends to light's speed but can not yet reach it. Besides, acceleration and time are finite. When a particle moves along the positive direction of radius vector, as long as it has a speed at the point $r$ 
\begin{equation}
V\geq{c}\sqrt{{\alpha\over{r}}}(1+{\alpha\over{r}})^{-{1\over{2}}}
\end{equation}
the particle can escape the gravitational field and has a speed $V\geq{0}$ when it reach the point $r\rightarrow\infty$ .
\par
    As for a photon, when it falls down the gravitational field in flat space-time, according to Eq.(51), its velocity and acceleration are
\begin{equation}
V={dr\over{dt_0}}=-c(1+{\alpha\over{r}})^{-1}
\end{equation}
\begin{equation}
\alpha={dV\over{dt_0}}={c^2\alpha\over{r^2}}(1+{\alpha\over{r}})^{-3}
\end{equation}
It can be seen that the photon is acted by repulsion and dose the retarded motion. When $t=0$ £¬let $r=r_0$ ,we have
\begin{equation}
ct_0=r-r_0+\alpha{ln}{r_0\over{r}}
\end{equation}
There is no any singularity in the area $r>\alpha$ . When the photon arrives at the point $r=0$, its speed $V=0$ £¬acceleration is also finite
\begin{equation}
\alpha=\lim_{x\rightarrow\infty}{c^2x^2\over{\alpha(1+x)^3}}\rightarrow{0}~~~~~~(x={\alpha\over{r}})
\end{equation}
But it takes photons an infinite time to reach the point $r=0$ .
\par
Therefore after the Schwazschild solution is transformed into flat space-time to describe, all original singularities disappear (In flat space-time, singularity appears in the form of over light speed's motion.). So it is obvious that the singularities in the general theory of relativity are actually caused by describing the theory in the curved space-time. The gravitational field itself has no singularities.
Meanwhile, it is known a photon can escape from the gravitational field when it moves along the direction of radius by the action of repulsion as long as it is not at the point $r=0$ . In this way, the black holes, at least the singular black holes with infinite densities and infinite small volumes, do not exist.
\par
If observers are in the reference system which falls free down the gravitational field, in this case, the time is $\tau$ and the distance between the observers and the center mass is $r$ £¬according to Eq.£¨9£©, we have
\begin{equation}
V={dr\over{d\tau}}=-\sqrt{{\alpha\over{r}}}=-\sqrt{{2GM\over{r}}}
\end{equation}
\begin{equation}
\alpha={dV\over{d\tau}}={dV\over{dr}}{dr\over{d\tau}}=-{\alpha\over{2r^2}}=-{GM\over{r^2}}
\end{equation}
\begin{equation}
\tau=-\int\sqrt{{r\over{\alpha}}}dr=-{2\over{3\sqrt{\alpha}}}(r^{3\over{2}}-r_0^{3\over{2}})
\end{equation}
They are just the results of the Newtonian gravitational theory. There are no singularities when $r>0$ . But when $r<GM/c^2r$ , the relative speed is over light's speed. When $r\rightarrow{0}$ , the relative speed becomes infinite. So the reference system falling free down the gravitational field is not yet a good reference system for the discussion of gravitational problems. The reason will be discuss in the third chapter.
\par
In brief, at least for the spherical symmetry gravitational field, it is more rational to study gravitational problems in flat space-time than in curved space-time. By transforming the solution of gravitational field equation to discuss in flat space-time, the problems can become more rational and simple. Some puzzling problems just as the singularity problems of black holes, the flat problem of the universal early stage and so on would be expounded. So it is necessary for us to re-examine the conclusions of the current general relativity by transforming them into flat space-time to study in order to get more rational results.
\par
\begin{center}
{\Large 2. The transformations in the general situations}\\
\end{center}
\par
Now let's generally prove that it is possible to transform the gravitational theories described in the curved
space-time into in the flat space-time. In the following discussion, the indexes of Egyptian letters are used to represent the 4-diamention quantities and the indexes of Latin letters represent the 3-diamention quantities. Let $x^{\alpha}$ represent the 4-diamention coordinates in curved space-time and $x^{\alpha}_0$ represent the 4-diamention coordinates in flat space-time. The 4-diamention linear elements in both space-times are individually
\begin{equation}
ds^2=dx^{\alpha}_0{d}x^{\alpha}_0=(dx^0_0)^2-dx^i_0{d}x^i _0
\end{equation}
\begin{equation}
ds^2=g_{\alpha\beta}{d}x^{\alpha}dx^{\beta}=g_{00}(dx^0)^2-2g_{0i}{d}x^0dx^i-g_{ij}{d}x^idx^j
\end{equation}
In the formulas, $x^0$ and $x^0_0$ are the time components. The equation of geodetic line of a particle moving in a gravitational field is
\begin{equation}
{d^2x^{\alpha}\over{ds^2}}+\Gamma^{\alpha}_{\beta\sigma}{dx^{\beta}\over{ds}}{dx^{\sigma}\over{ds}}=0
\end{equation}
For a certain gravitational field, suppose the metric tensor $g_{\alpha\beta}$  has been obtained by solving the Einstein's equation of gravitation field or other equations based on curved space-time, we can get from the integrals of Eq.(102)
\begin{equation}
x^i=x^i(s)
\end{equation}
as well as
\begin{equation}
x^0=x^0(s)
\end{equation}
From Eq.(104), we can obtain
\begin{equation}
s=s(x^0)
\end{equation}
Put it into Eq.(103), we get 
\begin{equation}
x^i(s(x^0))=x^i(x^0)=x^i(t)
\end{equation}
The particle's velocity and acceleration are
\begin{equation}
{dx^i(t)\over{dt}}=V^i(t)
\end{equation}
\begin{equation}
{d^2x^i(t)\over{dt^2}}={dV^i(t)\over{dt}}={\alpha}^i(t)
\end{equation}
In addition, two independent equations can be obtained by eliminating time t in the three equations of (106)
\begin{equation}
\phi_1(x^1,x^2)=0 ~~~~~~~~~~~~~\phi_2(x^1,x^3)=0
\end{equation}
If $x^i$ are the coordinates in the Euclidean space, $\phi_1$ and $\phi_2$ represent the equations of two columnar surfaces with the axial lines meeting at right angles. The intersecting line of the two columnar surfaces determined by Eq.(109) presents the orbit of a particle moving in the 3-dimention Euclidean space. If $x^i$ are the non-Euclidean space coordinates, $\phi_1$ and $\phi_2$ represent the two 2-dimention curved surface equations in the non-Euclidean space. Their intersecting line also represents the orbit of a particle moving in the 3-dimention non-Euclidean space 
\par
    On the other hand, as we know, any point on the 2-diamention non-Euclidean curved surface can find a one-to-one point in the 3-diamention Euclidean space, that is to say, the 2-dimention non-Euclidean curved surface can be inlaid into the 3-dimention Euclidean space. Therefore, any point at the interesting line of the two 2-dimention non-Euclidean curved surfaces can also find a one-to-one point in the 3-diamention flat Euclidean space. So we have the transformation relation between the points of geodetic line in the non-Euclidean space and the points in the Euclidean space
\begin{equation}
x^i=x^i(x^j_0)~~~~~~~~or~~~~~~~~~x^i_0=x^i_0(x^j)
\end{equation}
In order to get transformation relation between time $t$ and $t_0$ , by the condition $ds^2=constant$ , we have 
\begin{equation}
ds^2=(dx^0_0)^2-dx^i_0{d}x^i_0=g_{00}(dx^0)^2-2g_{oi}{d}x^0dx^i-g_{ij}{d}x^idx^j
\end{equation}
or
\begin{equation}
c^2dt^2_0=c^2g_{00}{d}t^2-2cg_{0i}{d}tdx^i-g_{ij}{d}x^idx^j+dx^i_0{d}x^i_0
\end{equation}
By considering Eq.(106) and (110), each item on the right side of Eq.(112) can be expressed as the function of time t, so we get transformation relation of time
\begin{equation}
t_0=\int\sqrt{g_{00}-2g_{0i}{dx^i\over{cdt}}-g_{ij}{dx^i\over{cdt}}{dx^j\over{cdt}}+{\partial{x^i_0}\over{\partial{x^l}}}{\partial{x^i_0}\over{\partial{x^k}}}{dx^l\over{cdt}}{dx^k\over{cdt}}}dt
\end{equation}
i.e.,
\begin{equation}
t_0=t_0(t)~~~~~~~~~~~~or~~~~~~~~~~~~~~t=t(t_0)
\end{equation}
     After the transformation relations of space-time coordinates are obtained, the equation (107) in the non-Euclidean space can be transformed into that in the Euclidean space. From Eq.(107),(110) and (114), we have 
\begin{equation}
{\partial{x^i}\over{\partial{x}^j_0}}{dx^j_0\over{dt_0}}={dx^i\over{dt}}=V^i(t)=V^i(t(t_0))=V^i(t_0)
\end{equation}
This is an equation set of three variables and one order about $dx^j_0/dt_0$ . We can obtain the velocity and acceleration in flat space by solving the equation set
\begin{equation}
{dx^i_0\over{dt_0}}=V^i_0(t_0)
\end{equation}
\begin{equation}
{d^2x^i_0\over{dt^2_0}}={dV^i_0(t_0)\over{dt_0}}=a^i_0(t_0)
\end{equation}
For the particle with static mass $m$ , the momentum in flat space is
\begin{equation}
\vec{P}_0={m\vec{V}_0\over{\sqrt{1-V^2_0/c^2}}}
\end{equation}
We get 
\begin{equation}
{d\vec{p}_0\over{dt}}={m\over{\sqrt{1-V^2_0/c^2}}}{d\vec{V}_0\over{dt}}+{m(\vec{V}_0\cdot\vec{\alpha}_0)\vec{V}_0\over{c^2(1-V^2_0/c^2)^{3/2}}}=\vec{F}
\end{equation}
Put Eq.£¨116£©and£¨117£©into Eq.(119), we get the dynamic equation and force of the particle in the gravitational field in flat space. In general, they are different from the Newtonian theory.
\par
As for photon, we can obtain the corresponding equations (116) and (117) from its geometric equation in curved space, then define photon's momentum in the same form of Eq.(56). After that, the dynamic equation of photon can be established in flat space-time. But it is unnecessary for us to discuss it nay more here. In this way, we have achieved the transformation of gravitation's descriptions from curved space-time to flat space-time.
\par
In the same, the gravitation's descriptions can also be transformed from flat space-time into curved space-time. From Eq.(119) in flat space-time, we can get by solving the equation
\begin{equation}
x^i_0=x^i_0(t_0)
\end{equation}
By introducing arbitrary coordinate transformation
\begin{equation}
x^\alpha_0=x^\alpha_0(x^\beta)
\end{equation}
we have
\begin{equation}
{dx^\alpha_0\over{ds^2}}={\partial{x}^\alpha_0\over{\partial{x}^\beta}}{dx^\beta\over{ds}}
\end{equation}
\begin{equation}
{d^2x^\alpha_o\over{ds^2}}={\partial^2x^\alpha_0\over{\partial{x}^\beta\partial{x}^\sigma}}{dx^\beta\over{ds}}{dx^\sigma\over{ds}}+{\partial{x}^\alpha_0\over{\partial{x}^\beta}}{d^2x^\beta\over{ds^2}}
\end{equation}
Using Eq.(116) and (121), we get from Eq.(101)
\begin{equation}
ds=(c^2-{dx^i_0\over{dt_0}}{dx^i_0\over{dt_0}})^{1\over{2}}dt_0=(c^2-V^i_0{V}^i_0)^{1\over{2}}dt_0=A(t_0(x^\beta))dt_0=A(x^\beta)dt_0
\end{equation}
Put it into Eq.(122), use Eq.(116) and (121), we have
\begin{equation}
{dx^\alpha_0\over{dt_0}}=A^{-1}{\partial{x}^\alpha_0\over{\partial{x}^\beta}}{dx^\beta\over{ds}}=V^\alpha_0(t_0(x^\beta))=V^\alpha_0(x^\beta)
\end{equation}
Here $V^0_0=c$. Eq.(125) is the equation set of four variables and one order about $dx^\beta/ds$ . We can obtain from Eq.(125)
\begin{equation}
{dx^\alpha\over{ds}}=B_\alpha(x^\beta)
\end{equation}
Put Eq.(124) into Eq.(123) and using Eq.(126), we get
\begin{equation}
A^-{2}{d^2x^\alpha_0\over{dt^2_0}}=B_\beta{B}_\sigma{\partial^2x^\alpha_0\over{\partial{x}^\beta\partial{x}^\sigma}}+{\partial{x}^\alpha_0\over{\partial{x}^\beta}}{d^2x^\beta\over{ds^2}}-A^{-1}{dA^{-1}\over{dt_0}}{dx^\alpha_0\over{dt_0}}
\end{equation}
By using Eq.(124)and(126) again, we have
\begin{equation}
{dA^{-1}\over{dt_0}}={\partial{A}^{-1}\over{\partial{x}^\beta}}{dx^\beta\over{dt_0}}={\partial{A}^{-1}\over{\partial{x}^\beta}}{dx^\beta\over{ds}}{ds\over{dt_0}}={\partial{A}^{-1}\over{\partial{x}^\beta}}B_\beta{A}
\end{equation}
We can write $\alpha^i_0(t_0)=\alpha^i_0(t_0(x^\beta))$ and have $\alpha^0_0=0$ in Eq.(117). By considering Eq.(125) and (128), Eq.(127) can be written as:
\begin{equation}
{\partial{x}^\alpha_0\over{\partial{x}^\beta}}{d^2x^\beta\over{ds^2}}=(A^{-2}F^i_0+B_\beta{V}^i_0{\partial{A}^{-1}\over{\partial{x}^\beta}}-B_\beta{B}_\sigma{\partial^2x^\alpha_0\over{\partial{x}^\beta\partial{x}^\sigma}})
\end{equation}
It is an equation set of four variables and one order about $d^2x^\beta/ds^2$ , in which all coefficients are the function of $x^\beta$ , so we can obtain
\begin{equation}
{d^2x^\alpha\over{ds^2}}=K_\alpha(x^\beta)
\end{equation}
The formula can be re-written as
\begin{equation}
{d^2x^\alpha\over{ds^2}}-K_\alpha({dx^\beta\over{ds}}{dx^\sigma\over{ds}})^{-1}{dx^\beta\over{ds}}{dx^\sigma\over{ds}}={d^2x^\alpha\over{ds^2}}-K_\alpha{B}^{-1}_\beta{B}^{-1}_\sigma{dx^\beta\over{ds}}{dx^\sigma\over{ds}}=0
\end{equation}
Let
\begin{equation}
\Gamma^\alpha_{\beta\sigma}=-K_\alpha{B}^{-1}_\beta{B}^{-1}_\sigma
\end{equation}
Eq.(131) becomes
\begin{equation}
{d^2x^\alpha\over{ds^2}}+\Gamma^\alpha_{\beta\sigma}{dx^\beta\over{ds}}{dx^\sigma\over{ds}}=0
\end{equation}
Regarding $\Gamma^\alpha_{\beta\sigma}$ as the Christoffel sign, Eq.(133) is just the geodetic line equation in the new reference system. From the definition
\begin{equation}
\Gamma^\alpha_{\beta\sigma}={1\over{2}}g^{\alpha\beta}({\partial{g}_{\rho\sigma}\over{\partial{x}^\beta}}+{\partial{g}_{\rho\beta}\over{\partial{x}^\sigma}}-{\partial{g}_{\beta\sigma}\over{\partial{x}^\rho}})
\end{equation}
we known that the number of independent $\partial{g}_{\alpha\beta}/\partial{x}^\sigma$ is just the same as the number of independent $\Gamma^\alpha_{\beta\sigma}$ . Meanwhile, $g_{\alpha\beta}$ can be got from the following formula
\begin{equation}
g_{\alpha\beta}=\int{\partial{g}_{\alpha\beta}\over{\partial{x}^\sigma}}dx^\sigma
\end{equation}
So the metric tensors can be determined by Eq.(134) and (135). It is noted that if Eq.(121) is put into Eq.(101) directly, we have
\begin{equation}
ds^2=g'_{\alpha\beta}{d}x^\alpha{d}x^\beta
\end{equation}
                                                                  £¨136£©
The metric tensor $g_{\alpha\beta}$ is different from that shown in Eq.(135) in general. The metric tensor $ g'_{\alpha\beta}$ shown in Eq.(136) is the Euclidean metric in essence for it can return to the original form Eq.(101) by an inverse transformation. But the metric tensor $g_{\alpha\beta}$  determined by Eq.(134) and (135) can not in general, so they are the non-Euclidean metrics in general.
\par
Up to now we have achieved the transformation of gravitation's descriptions between the curved space-time and the flat space-time, and proved their equivalence. The difference is that in curved space-time particles move along the geodetic lines without forces acting on them, but in flat space-time particles move along the non-geodetic lines acted by gravitation. What kinds of descriptive forms are taken depends on convenience in principle, but as shown above, the practical results should be considered.\\
\begin{center}
{\Large 3. Discussions on some foundational concepts}\\
\end{center}
\par
    1. Is the space-time curved or flat after all when gravitational field exists?
\par
This is first a problem of measurement. Whether can we answer this problem by the direct measurement? The answer is negative. Even thought the space-time is curved when the gravitational field exists, we can not detect it by the direct measurement. This is owing to the fact that before the measurement we have to define standard ruler and clock. But only in flat space-time, we can do them. In curved space-time, we have no definitions of standard ruler and clock. When the ruler and clock defined in flat space-time are put into gravitational fields, they would change or become curved synchronously with the fields, so that the measurements can not show the changes of curved level of space-time. The ruler and clock in the gravitational fields can not free themselves from the effects of gravitational fields, so it is impossible to show whether space-time is curved or flat when gravitational field exists by the direct measurement.
\par
What we can do is to use indirect methods, for example, to observe the orbits of test particles orbit or the red shifts of spectral lines in the gravitational fields to decide the curved level of space-time. However, as shown above, we can describe the orbit of test particle in gravitational field by either geometric equation in curved space-time, or the dynamic equation in flat space-time. We can also explain the red shifts of spectral lines in the gravitational fields as the results of time delay or the potential energy's changes. Two methods are equal to each other. So it is obvious that space-time itself can not be designated as curved or flat actually. The reality is what kind of reference systems, curved or flat, we choose to describe it. If the curved reference system is chosen, the space-time is curved. If the flat reference system is chosen, the pace-time becomes flat. It is meaningless to talk about space-time itself curved or flat. So we should only use the concept of curved or flat reference system, in spite of the concept of curved or flat space-time.
\par
    2. The equivalent principle 
\par
According to the weak equivalent principle, gravitational mass and inertial mass are equivalent to each other. Let $m$ represent static mass,  represent inertial motion mass, we have
\begin{equation}
m_1={m\over{\sqrt{1-V^2/c^2}}}
\end{equation}
Let $m_G$ represent gravitational mass, comparing Eq.(22) with the Newtonian formula of gravitation and considering the relation $\vec{L}=\vec{V}\times\vec{r}$ , we get
\begin{equation}
m_G=m\sqrt{1-{V^2\over{c^2}}}[1+{3(\vec{V}\times\vec{n})^2\over{c^2}}]
\end{equation}
Here $\vec{n}$ is the unit radius vector. It is obvious $m_1\neq{m}_G$ in general situations. Only when $V=0$ , they are equal to each other. In fact, all completed Eotvos experiments only prove that gravitational mass and inertial mass are equal to each other when the testing bodies on the two ends of cantilever beam are at rest each $other^{(4)}$. It has not yet be proved that they are equivalent when there exists relative motion between them. It should be noted that the formula (138) is the result of Einstein's field equation, showing that gravitational mass and inertial mass are not equivalent actually when the factor of speed is considered.
\par
3. The principle of general relativity
\par
The paper's conclusions are completely based on the Einstein equation of gravitational field, no any new hypothesis is introduced besides transforming the theory to the flat reference system to discuss. According to the general theory of relativity, it is equivalent to discuss physical problems in any reference system in nature. No one is more superior. Since we can discuss gravitational problems in any reference system, we can also discuss them in flat reference system. However, the results show that flat reference system seems more superior for the discussion of gravitational problems. The results contradict the principle of general relativity. So we have to discuss this problem further.
\par
Einstein established special relativity that denied the existence of absolutely static reference system. Later, he put forwards the principle of general relativity trying to cancel the superior position of the inertial reference systems. If we consider the principle of general relativity as that the motion equations are covariant, or the basic forms of the motion equations are the same in any reference system, the principle is all right. However, it does not mean that the concrete forms of motion equations and their solutions are the same. In special relativity, the 4-diamention coordinate transformation means that the relative speed is introduced. Let $p_\mu$ represents the 4-diamention momentum,
 $F_\mu$ represents the 4-diamention force. The basic form of motion equation $dp_\mu/dt=F_\mu$ is unchanged when an inertial reference system is transformed into another inertial reference system moving in a uniform speed.. But the concrete forms of the 4-diamention force $F_\mu$ and particle's motion, as well as and some physical quantities would change. For example, length constrict, time delay, moving mass increasing and the form of force changing and so on, though according to special relativity, these changes only have relative meanings.
\par
 In general relativity, the coordinate transformations involve more problems. At present, it is considered that a solution of gravitational field equation can still represent the same field after the solution has been transformed into another new reference system. This conclusion is worthy of further discussion and consideration. If the coordinate transformation is carried out in the 3-diamention space, there is no any problem. But in the 4-diamention space-time, because time is involved, the situation is completely different. In the general theory of relativity, the 4-diamention coordinate transformation means that acceleration or non-inertial reference system is introduced. According to the principle 
of equivalence, non-inertial reference system is equal to gravitational field. The transformation from one non-inertial reference system to another means that a gravitational field is changed to another. So the coordinate transformations would change physical processes. Speaking clearly, for a gravitational field with a determinate form, suppose we have obtained the solution by solving the Einstein's field equation, if the solution is transformed into another reference system, the form of the solution would change, though the basic form 
of field equation is unchanged. In the light of the principle of equivalence, it means that a new gravitational field is introduced and the original solution loses its meaning. Therefore, a determinate gravitational field can only corresponds to a determinate metric, arbitrary coordinate transformation is forbidden according to the principle of equivalence. Unless the same results can be reached in new reference system for all problems, but this is impossible in general. For example, we can not calculate the perihelion precession of Mercury and other experiments and get the same results in the Edington or Kruskal coordinate systems. This is just the reason why the energy of gravitational field can not
 be defined well in the current theory. For a definite gravitational field with a certain of symmetry, we can only define its energy in the definite metric with the same symmetry. But it is allowed to transform the solutions into the inertial or flat reference systems to discuss. In this case, what is done is to transform the geodesic lines into the dynamic equations of gravitation without any attached force being introduced. Or speaking more clearly, there exist no relativity and arbitrariness in the description of gravitation. A certain of absoluteness is needed for us to describe gravitation. We should establish a united standard for gravitation. Only based on the flat reference system, we can do it.
\par
    It can be seen from discussions above that though the Einstein's theory of gravitation has obtained great succession, there still exist some problems in its theoretical and logical foundation which need to be cleared and reformed so that the theory can become more rational. It is obvious that we need renewing some ideas about the essence of space-time and gravitation. The author will discuss them in detail later.

\end{document}